\documentclass[]{spie}  

\usepackage{amsmath,amsfonts,amssymb}
\usepackage{graphicx}
\usepackage{soul}
\usepackage[colorlinks=true, allcolors=blue]{hyperref}
\usepackage{outlines}
\usepackage{booktabs}
\usepackage{varwidth}
\usepackage{floatrow}
\usepackage{algorithm}
\usepackage[noend]{algpseudocode}
\usepackage{multirow}
\usepackage{subcaption}
\usepackage[normalem]{ulem}
\useunder{\uline}{\ul}{}

\title{ Direct Optimisation of {\LARGE$\boldsymbol\lambda$} for HDR Content Adaptive Transcoding in AV1}

\author[a]{Vibhoothi*}
\author[a]{François Pitié}
\author[a]{Angeliki Katsenou}
\author[a]{Daniel Joseph Ringis}
\author[b]{Yeping Su}
\author[b]{Neil Birkbeck}
\author[b]{Jessie Lin}
\author[b]{Balu Adsumilli}
\author[a]{Anil Kokaram*}

\affil[a]{Trinity College Dublin, Dublin, Ireland}
\affil[b]{Google Inc., CA, USA}

\authorinfo{* vibhoothi@tcd.ie,  anil.kokaram@tcd.ie}

\begin{document} 
\maketitle

\begin{abstract}
Since the adoption of VP9 by Netflix in 2016, royalty-free coding standards continued to gain prominence through the activities of the AOMedia consortium. AV1\cite{av1paper}, the latest open source standard, is now widely supported. In the early years after standardisation, HDR video tends to be under served in open source encoders for a variety of reasons including the relatively small amount of true HDR content being broadcast and the challenges in RD optimisation with that material. AV1 codec optimisation has been ongoing since 2020 including consideration of the computational load\cite{svtav1spie2021}. In this paper, we explore the idea of direct optimisation of the Lagrangian $\lambda$ parameter used in the rate control of the encoders to estimate the optimal Rate-Distortion trade-off achievable for a High Dynamic Range signalled video clip. We show that by adjusting the Lagrange multiplier in the RD optimisation process on a frame-hierarchy basis, we are able to increase the Bjontegaard difference\cite{bdrate} rate gains by more than 3.98$\times$ on average without visually affecting the quality.


\end{abstract}


\section{Introduction}
\label{sec:intro}  

In recent years, the growth in delivery of video at scale for broadcast and streaming applications (from Netflix, YouTube, Disney etc.) has inspired further research into content-adaptive transcoding. The goal is to deliver high-quality content at progressively lower bitrates by adapting the transcoder for each input at a fine-grained level of control. In 2013, YouTube was the first to adopt this strategy for its User-Generated-Content (UGC) by building a pipeline that is based on clip popularity by re-processing a clip with an enhanced pre-processor in combination with a different built-in transcoder. Around the same time, Netflix's seminal work on per-clip and per-shot encoding~\cite{netflix} for High Value Content (HVC) videos showed that an exhaustive search of the coding parameter space can lead to significant gains in Rate-Distortion (RD) tradeoffs per clip. These gains offset the high one-time computational cost of encoding as the same encoded clip may be streamed to millions of viewers across many different Content Delivery Networks (CDNs)\cite{renznik_internetvideo2001}, thus effectively saving bandwidth and network resources. That idea has since been revisited  and has become more efficient by applying the Viterbi algorithm across shots and parameter spaces~\cite{katsavounidis2018video}. Over the past years a lot of researchers have focused on the optimisation of a high-level parameter (target bitrate or quantisation factor or objective quality) to generate an optimal bitrate ladder for a clip as part of a Adaptive Bitrate Streaming (ABR)\cite{reznik2018abrstreaming, TimmererSurvey, KatsenouOJSP2021}.

In our previous work~\cite{pcs2021ringis}, we showed that the RD tradeoff can be directly addressed by applying a numerical optimisation scheme to estimate the appropriate Lagrangian $\lambda$ multiplier for a given clip (see Section~\ref{sec:background}) for standard dynamic range (SDR) videos. We observed an average BD-rate improvement of 1.9\% for HEVC, 1.3\% for VP9~\cite{SPIERingis}, and 0.5\% in AV1~\cite{icip2022paper}. In our latest work~\cite{icip2022paper}, we further demonstrated that additional BD-rate(\%) gains, from 0.5\% to 4.9\%, for AV1 could be achieved by adopting a per frame-type optimisation. 

In this paper, we explore the idea of $\lambda$ optimisation on High-Dynamic Range (HDR)/Wide Color Gamut (WCG) material\cite{itu_hdr}. HDR/WCG systems can capture, process, and reproduce a scene conveying the full range of perceptible shadow and highlight details beyond normal dynamic range (SDR) video systems.

Similarly to our latest work~\cite{icip2022paper}, we propose a content-adaptive transcoder optimisation at a global and a deeper frame-type level. The core new ideas are i) consideration of  RD Optimisation (RDO) for HDR content, ii) optimisation of the RDO parameters on a frame-hierarchy basis, and iii) investigation of various convergence criteria that result in the minimisation of the computational load.
Our experiments in Section~\ref{sec:expt-results} demonstrate that a frame-based tuning of video encoder can lead to an average gain of 1.63\% of BD-rate (best recorded gain of 9.3\%) compared to the standardised method. Moreover, the average gains per shot range between 0.58 and 3.43\% for HDR video content in AV1.

Section~\ref{sec:background}  gives an overview of previous research work and $\lambda$ definition in rate control. Section \ref{sec:directopt} explains the proposed methodology as well as the multi-dimensional optimisation. Section \ref{sec:expt-setup} then details the experimental set-up including the test sequences, the keyframe combination selection,  the framework implementation. Section \ref{sec:expt-results} reports on the experimental findings.

\section{Background}
\label{sec:background}

The work of Sullivan et al~\cite{sullivan1998rate} laid the foundations for optimising the RD tradeoff in modern video codecs. By taking a Lagrange multiplier approach the joint optimisation problem is posed as the minimisation of $J = D + \lambda R$, where $\lambda$ is the Lagrange multiplier controlling the tradeoff. This idea is the basis of the RDO process used especially in making mode decisions in modern codecs. The independent variable in this optimisation is usually $qp$, a quantiser step size. Increasing $qp$ reduces rate $R$ but increases distortion $D$. Also, a different choice of $\lambda$ yields different $R,D$ pairs. 

Different codecs devised different recipes to derive the optimal $\lambda$ value for RDO through an empirical relationship with $qp$. In libaom-AV1~\cite{av1paper} $\lambda$ is empirically related to $q_i$ (Quantizer Index: $\approx qp*4$ in the AV1 codebase), as follows:
\begin{equation}
  \lambda = q_{dc} ^2 \times (A + 0.0035 \times q_{i}), \label{lambdaeq:av1}
\end{equation}
where $A$ is a constant depending on the frame type ($3.2 \leq A \leq 3.3$) and $q_{dc}=f(q_i, A)$ is defined through a discrete valued Look Up Table (LUT) ($0 \leq q_i \leq 255$ for AV1). 
This $\lambda-qp$ relationship is not necessarily optimal for a particular clip because the empirical relationship was derived for optimality over an entire test corpus. To maximise gains, $\lambda$ should be content dependent. 



{\noindent \bf{Per Clip \texorpdfstring{$\lambda$}{lambda} Optimisation}}. The idea of adapting $\lambda$ based on video content is not entirely new. Zhang and Bull~\cite{zhangbulllambdahevc} altered $\lambda$ based on
distortion statistics on a frame-basis for HEVC. In our previous work~\cite{EIRingis,SPIERingis, pcs2021ringis}, we introduced the idea of an adaptive $\lambda$ on a clip basis, using a single modified $\lambda = k\lambda_o$ across all the frames in a clip. Here, $\lambda_o$ represents the default value deployed from the relevant empirical relationship e.g. Eq.~(\ref{lambdaeq:av1}). 
In order to find the optimal $\lambda$ value, we deployed numerical optimisers (Brent's method and Golden-search~\cite{numericalmethods}) that minimised the BD-rate  as a cost function. Later work~\cite{pcs2021ringis}, considered the use of Machine Learning techniques to reduce the required computational load.

{\noindent \bf{Per Clip, Per Frame-Type \texorpdfstring{$\lambda$}{lambda} Optimisation}}. In our latest work~\cite{icip2022paper}, we showed that this method of global $\lambda$ tuning yields average BD-rate(\%) gains of only 0.539\% and 0.097\% for AV1 and HEVC, respectively. These modest improvements are probably due to the fact that current modern video encoders are content-adaptive by nature and include many new improvements such as partition tools, Inter/Intra prediction tools and modern hierarchical reference frame-structure. Another important aspect to consider is that the heuristics and empirical shortcuts used in this content-adaptive implementation of the Lagrangian parameter, deviates from classical RDO theory, which normally requires $\lambda$ to be constant across the sequences over which distortion is measured. For example in Eq.(\ref{lambdaeq:av1}) $\lambda$ changes with frame type. 
Motivated by this observation, in our latest work~\cite{icip2022paper}, we studied the effect of isolating the optimisation of $\lambda$ for different frame types. Results on SDR sequences showed that optimising $\lambda$ purely for Keyframes (KF), Golden-Frames (GF), Alternate-reference Frames (ARF), leads to average BD-rate gains of 4.92\% compared to global $\lambda$ optimisation (only 0.54\% BR-Rate gains).

{\noindent \bf{ HDR in AV1}}. The quality of compressed 4K and 8K HDR content was evaluated by Pourazad et. al.~\cite{hdr4kcomparision}, drone content from P. Topiwala et.al.~\cite{topiwala2021hdr}, and gaming (Nabajeet et. al.~\cite{barman_ugc_hdr}).
More relevant here is the work of Zhou et. al~\cite{zhou_lambda_hdr} for HEVC that expresses distortion $D$ in terms of the HDR-VDP-2 quality metric~\cite{hdrvdp2}. They presented an algorithm for prediction of $\lambda$ at the CTU level which resulted in 5\% BD-rate improvement w.r.t. a reference implementation of HEVC (HM16.19).

\section{Direct \texorpdfstring{$\lambda$}{lambda} Optimisation in AV1}
\label{sec:directopt}

As noted in Section \ref{sec:background}, the encoder is determining $\lambda$ independently on a frame basis inside the encoder. In this work, we explore the impact of treating $\lambda$ optimisation as a multi variable search problem at a frame basis w.r.t. HDR content. The following sections expand our main strategies for the direct optimisation of the $\lambda$ parameter.

{\noindent \bf{BD-rate Optimisation}}. For the $n^{\text{th}}$ RDO decision in clip $m$, we propose that $\lambda_n=k_n \lambda_0$.  We estimate ${\bf k} = [k_1,\dots,k_N]$ (where we assume $N$ RDO decisions) to maximise the BD-rate gain using MS-SSIM~\cite{msssimpaper} as the quality metric ($Q_m$). The cost function $C_m({\bf k})$ can therefore be formulated as:
\begin{align}
    C_m({\bf k}) & = \text{BD-rate}(R_m({\bf k}\lambda_o,  Q),R_m(\lambda_o, Q)) \nonumber \\
    & \propto \int_{Q_1}^{Q_2} \left(R_m({\bf k}\lambda_o, Q)-R_m(\lambda_o, Q)\right) dQ \; ,
\label{eq:bdr_single}
\end{align}
where $R_m({\bf k}\lambda_o, Q)$ is the bitrate of the $m^{\text{th}}$ clip at quality $Q$, using ${\bf \lambda} = {\bf k}\lambda_o$ for the $N$ RDO decisions and $Q_1,Q_2$ defined as usual\cite{bdrate}. $R_m(\cdot,\cdot)$ is derived from the MS-SSIM-based RD curve  generated using $P$ $qp$ measurements.  Here we use $P=5$: $qp=$\{27, 39, 49, 59, 63\}.
\begin{algorithm}[t]
\caption{Proposed Optimisation Workflow}
\begin{algorithmic}[1]
\State Initialise video set, $qp$ points, codec and associated options. 
\State [Run-$A$] Encode each clip in parallel with $P$ threads, where $P$ is a number of $qp$ points. 
\State Compute RD-points and collect metrics using {\tt libvmaf}.
\State This establishes the RD curve corresponding to $\lambda_o$
\State [Run-$B$] Deploy selected Numeric Optimiser with initial condition ${\bf k} = 1$
\State On each successive iteration $i$ of the Optimiser Do 
\State \hskip 1em Encode the given clip using the new value of ${\bf k}={\bf k}_i$
\State \hskip 1em Compute the cost function $C_m({\bf k}_i)$ using reference RD curve for BDRATE calcuation from Run-$A$
\State Terminate optimiser using default convergence and stopping criteria.
\end{algorithmic}
\label{algorithm:blackbox}
\end{algorithm}

The flow of the optimisation framework is reported in Algorithm~\ref{algorithm:blackbox}. We can see that repeated computations of the BD-rate are required, and this incurs a huge computational cost. To address this, we deploy the idea of proxies for parameter selection as proposed by Ping H. et. al.\cite{ping_spie_proxy, ping_pcs_proxy}. They observe that using different speed settings of the encoder at the same target quality/quantizer level results primarily in bitrate differences that are directly proportional to the content complexity (see section \ref{expt-results:proxy}). That idea also extends to the use of lower resolution proxies. Therefore we can reduce computational load by performing optimisation using faster encoder presets {`em and} lower resolution proxies.

{\noindent \bf{Multi-Dimensional Optimiser}}. Our previous study for finding the optimal $\lambda$ multiplier in AV1 used Brent's line search method~\cite{icip2022paper}. The focus was on applying a modifier $k$ to only one sub-module in the encoder. However, here we explore the scenario of a multiple dimensional search for finding the optimal $\lambda$ for multiple frame types each associated with a different $k$. For this multi-dimensional search, we have various options like Nelder-Mead Simplex\cite{simplex_paper}, Conjugate Gradient\cite{nocedal2006conjugate}, Powell's method\cite{powellmethodpaper}, etc. 

In order to select a suitable optimiser, we conducted a simple analysis by carrying out an exhaustive grid-search on a single video to study the surface of our objective function. For this study, we chose the clip NocturneRoom from the AOM Common-Testing-Configuration (AOM-CTC)\cite{aomctc} set, and optimised $\lambda$ for two different frametypes: $\lambda_{KF}$ for the KFs and $\lambda_{GF/ARF}$, for the GF/ARFs. For all other frame-types, $\lambda$ was set to the default. The grid search range for  both $\lambda_{KF}$ and $\lambda_{GF/ARF}$ is 0.6 to 5.4, in steps of 0.1.
This provided 2,401 anchor points resulting in a total of 12,005 RD points for analysis. 

\begin{figure}
   \begin{center}
   \includegraphics[width=.5\linewidth]{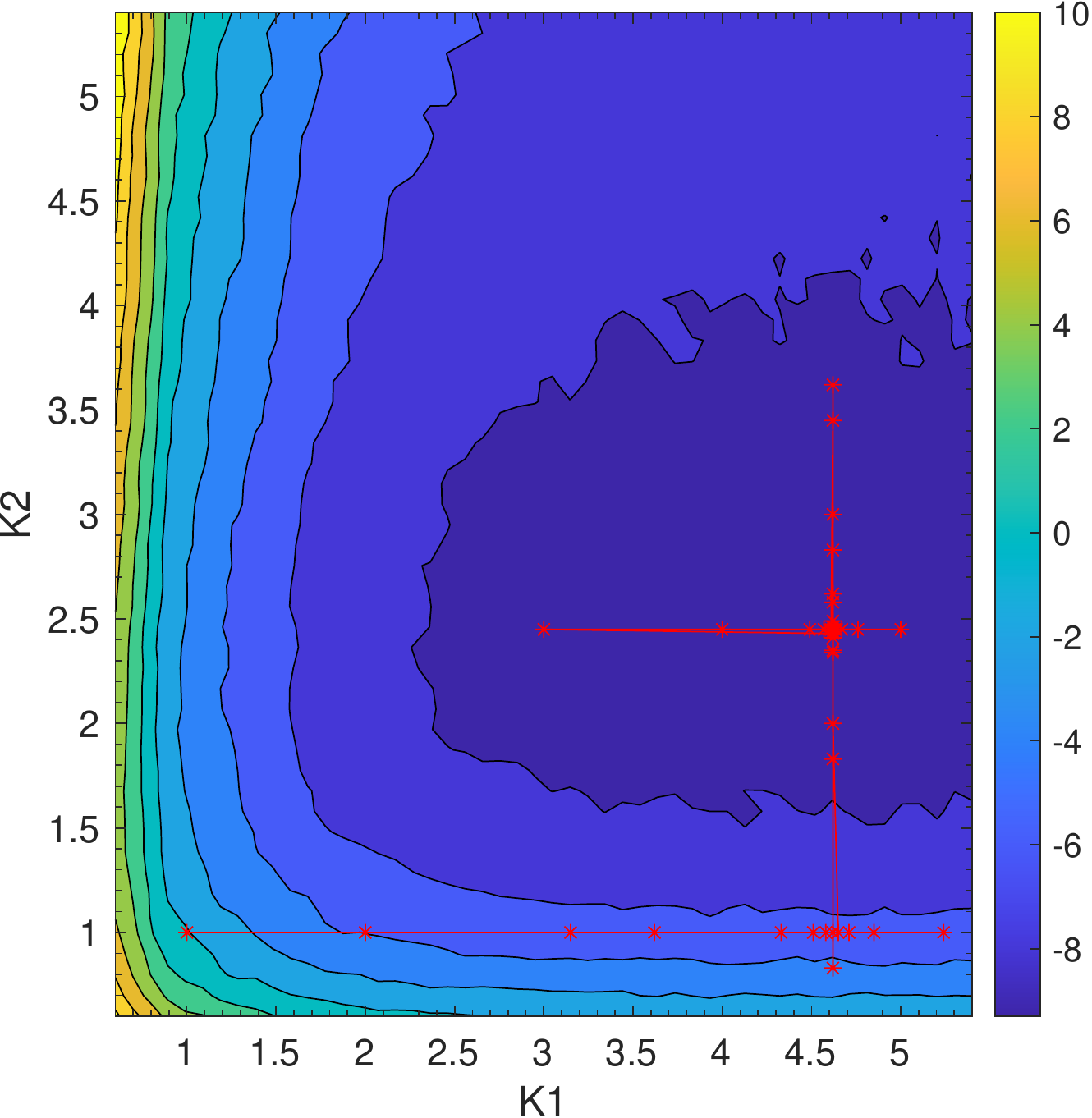}
   \end{center}
   \vspace{-2em}
   \caption{Plot of the optimisation surface, where X axis represents the $\lambda_{KF}$[K1] multiplier value and Y axis the $\lambda_{GF/ARF}$[K2] value, contour is $C_m(k_1,\dots,k_n)$. The red line shows the variation of K1 and K2 values when Powell's method~\cite{powellmethodpaper} is deployed for direct optimisation.}
    \label{fig:surface-plot} 
  \end{figure}

Figure \ref{fig:surface-plot} shows the contour plot of the BD-rate \% (MS-SSIM) objective function for a sample clip. The surface is clearly smooth, and a gradient based method is expected to converge to a sub-optimal solution (local minimum) due to very low gradient. This observation was confirmed by testing gradient based methods such as the Nelder-Mead Simplex method\cite{simplex_paper} and the Conjugate Gradient\cite{nocedal2006conjugate}. Both of these methods converged erroneously after the first iteration as the gradient was very close to zero. Therefore, we explored line-search constrained methods. One of the best performing was the modified Powell method~\cite{powellmethodpaper}, which succeeded to reach the global minima for the test clip (see red-lines on the Figure \ref{fig:surface-plot}).

\section{Experimental Setup}
\label{sec:expt-setup}
\subsection{4K HDR Corpus}
\label{sec:expt-setup:corpus}
\begin{figure}
\CenterFloatBoxes
\begin{floatrow}
\ffigbox
  {\includegraphics[width=.85\linewidth]{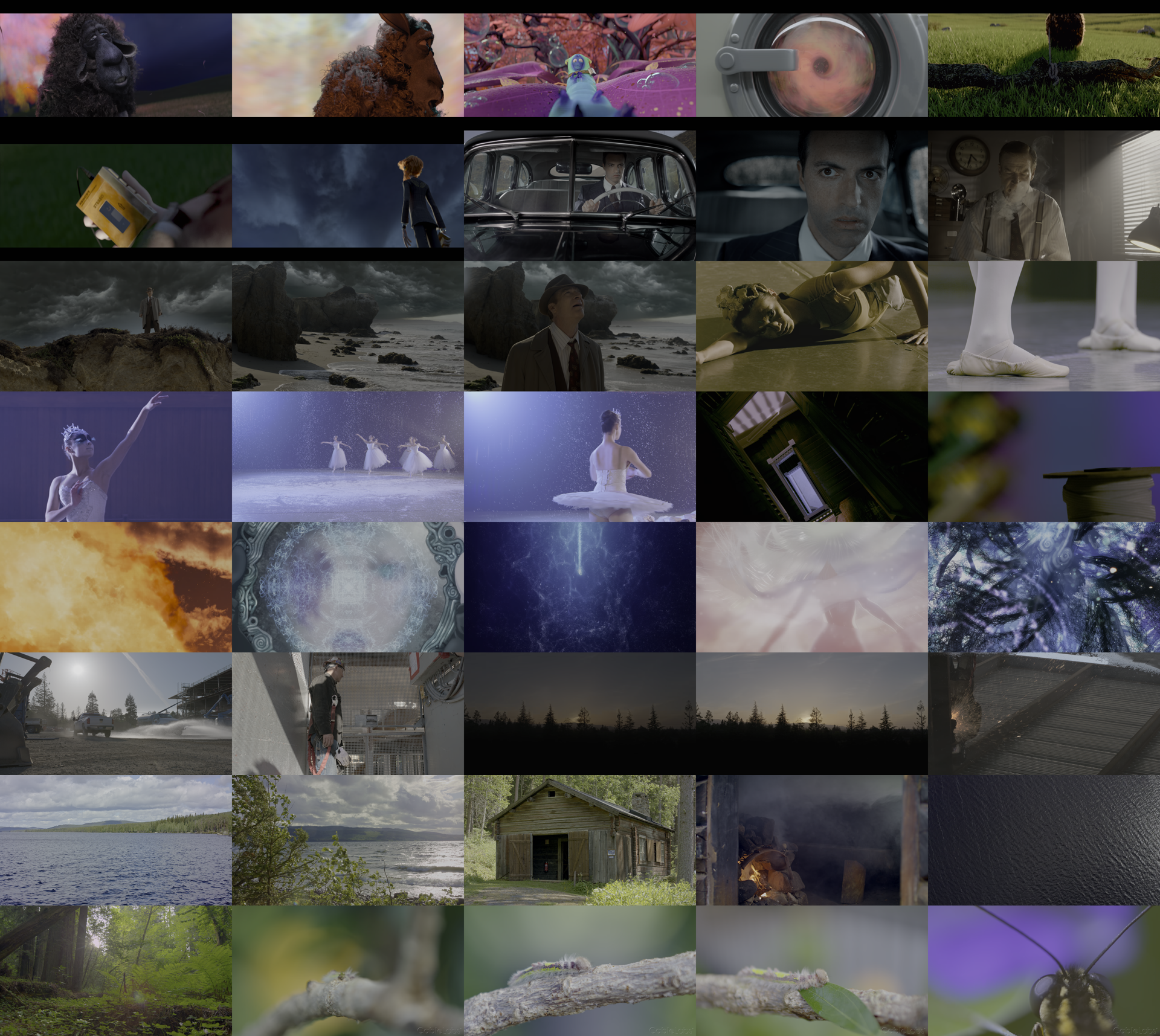}}
  {\caption{Sample frames from the corpus.}\label{fig:seq-grd}}
\killfloatstyle
\ttabbox
  {
  \small
  \begin{tabular}{@{}ll@{}}
   \toprule
   \textbf{Shot Group} & \textbf{Description} \\ \midrule
   Cosmos~\cite{netflixopencontent} & \begin{tabular}[c]{@{}l@{}} Vibrant animated sequence, high \\ temporal complexity, 24fps       \end{tabular} \\
   Meridian~\cite{netflixopencontent, aomctc} & \begin{tabular}[c]{@{}l@{}}Natural sequence, high \\ spatial complexity, 59.94fps\end{tabular}       \\
   Nocturne~\cite{netflixopencontent, aomctc} & \begin{tabular}[c]{@{}l@{}} Natural sequence, medium \\ spatial complexity, 60fps \end{tabular} \\
   Sol Levante\cite{netflixopencontent, aomctc} & \begin{tabular}[c]{@{}l@{}}Animated sequence, medium-high \\ temporal complexity, 24fps\end{tabular} \\
   Sparks~\cite{netflixopencontent, aomctc} & \begin{tabular}[c]{@{}l@{}}Sequence with medium motion\\ and wide dynamic range, 59.94fps\end{tabular} \\
   SVT\cite{svt2022} & \begin{tabular}[c]{@{}l@{}}Very high natural \\complexity sequences, 50fps \end{tabular} \\
   Cables 4K\cite{cablelabs}  & \begin{tabular}[c]{@{}l@{}}Outdoor sequences with\\ moderate complexity, 59.94fps\end{tabular} \\ 
   \bottomrule
   \end{tabular}
  }
  {\caption{High-level description of the shots.}\label{tab:seq-desc} }
\end{floatrow}
\end{figure}

For our experimental studies, we formed a video corpus consisting of 50 video clips (6500 frames) curated from various public sources. All the videos are normalized to BT.2020 color primaries with SMPTE2084 Perceptual Quantizer (PQ) transfer function and represented in the YUV colourspace inside YUV-Y4M containers. 
All the conversions and normalization of the clips are implemented with HDRTools~\cite{hdrtools}. The configuration file for the conversion to YUV Space with PQ Signal is available in our project page\footnote{Project page: \url{https://gitlab.com/mindfreeze/spie2022}}. The 50 clips contain 130 frames, a resolution of 3840x2160/4096x2160, and can be further grouped into 7 shot groups. Figure~\ref{fig:seq-grd} illustrates sample frames of the dataset and
Table~\ref{tab:seq-desc} gives a short description of the content of these 7 shot groups. 
More information on the dataset, including computed Spatial and Temporal Information (SI and TI)~\cite{itup910} and Dynamic Range (DR) can be found in our project page\footnotemark[\value{footnote}].

\subsection{Keyframe Selection}
\label{sec:expt-setup:kf-opt}
Reference frames (henceforth called keyframes) in AV1, typically contain $5$ to $10$ times more bits than other frames. Therefore, we target the optimisation of the bit allocation of these keyframes. In AV1, in order to code an Inter frame, references up to 8 keyframes~\cite{av1:adaptivepred} are used. The encoder chooses from multiple frames in both forward and backward direction~\cite{av1compound}.
For the simulations presented, we consider multiple combinations of the 3 keyframe types: the reference Intra-coded frame {\em KF}, an {\em ARF\_FRAME (ARF)} used in prediction but does not appear in the display, and an Inter coded frame which is coded at higher quality {\em GOLDEN\_FRAME (GF)}.

\subsection{Framework Implementation}
\label{sec:expt-setup:impl}
For the simulations, Random-Access (RA) encoding mode was chosen as per AOM-CTC~\cite{aomctc}. This mode is commonly used for streaming as it allows users to randomly seek into any frame of the clip. We deployed a stable release for AV1 ({\tt libaom-av1-3.2.0, 287164d}) with modifications to allow $k$ to propagate to the desired mode from a command-line argument. 

The objective metrics for Quality and Rate (RD measurements) at the selected $qp$ settings were computed using libvmaf~\cite{libvmafurl}, a standard open-source video quality evaluation library. Our software framework for performing these experiments with AV1 is based on AreWeCompressedYet~\cite{awcy}. 


\section{Experiments \& Results}
\label{sec:expt-results}
\subsection{Proxy Processing}
\label{expt-results:proxy}
As discussed earlier, optimisation requires many encodes, and we need to consider faster proxy presets to make these experiments practical. We have investigated three proxy modes for use during the optimisation:  
(4K S2-S2), the default non-proxy encode at the original 4K resolution using AV1 Speed 2 preset as in the final setting, (4K S2-S6), which encodes videos at 4K resolution using AV1 Speed 6 preset, and (1080p S2-S6), which operates at a (Lanczos 5) downsampled 1080p resolution with speed preset 6.
    
Table~\ref{tab:proxy-table} presents the BD-rate gains on the 4 sequences of the \textit{av2-g1-hdr-4k} set from AOM-CTC. The optimisation is performed in this case for a single global $k$ using Brent's method\cite{numericalmethods}. 
It is clear that using the proxy method (1080p S2-S6) reduces the encoding complexity by an average of 4.8$\times$, with negligible degradation in quality (BD-rate) when compared against full-resolution optimisation mode (4K S2-S2). We also note that BD-rate gains at (4K S2-S6) are almost identical to (1080p S2-S6), but with about 30\% slower encoding speed.
Given that processing time for our subsequent experiments takes in the order of 100's of hours, we therefore use this (1080p S2-S6) proxy setting in the rest of the study. Hence in results reported in Table~\ref{tab:full-gains-table} we use that proxy to estimate optimal values of $k_1,\dots, k_n$; then evaluate BD-rate gains using the original material i.e. 4K with S2 preset. We observe of course some differences between this approach and using 4K S2 throughout but it is the more pragmatic approach validated by our results in Table~\ref{tab:proxy-table}. 

\begin{table}
\caption{Proxy encoding time(hours), estimated $\lambda$ multiplier value, and BD-rate gains (\%) MS-SSIM for the different proxy encoding speeds and video resolutions.  }
\label{tab:proxy-table}
\scriptsize
\begin{tabular}{@{}lrrrrrrrrr@{}}
\toprule
\textbf{} & \multicolumn{3}{l}{\textbf{Encoding Time (hrs)}} & \multicolumn{3}{l}{\textbf{$\lambda$ Multiplier}} & \multicolumn{3}{l}{\textbf{BD-rates (\%)}} \\ \cmidrule(lr){2-4}\cmidrule(lr){5-7}\cmidrule(lr){8-10}
\begin{tabular}[c]{@{}l@{}}Proxy\\ Settings \end{tabular} & \multicolumn{1}{l}{\begin{tabular}[c]{@{}l@{}}4K\\ S2-S2\end{tabular}} & \multicolumn{1}{l}{\begin{tabular}[c]{@{}l@{}}4K\\ S2-S6\end{tabular}} & \multicolumn{1}{l}{\begin{tabular}[c]{@{}l@{}}1080p\\ S2-S6\end{tabular}} & \multicolumn{1}{l}{\begin{tabular}[c]{@{}l@{}}4K\\ S2-S2\end{tabular}} & \multicolumn{1}{l}{\begin{tabular}[c]{@{}l@{}}4K\\ S2-S6\end{tabular}} & \multicolumn{1}{l}{\begin{tabular}[c]{@{}l@{}}1080p\\ S2-S6\end{tabular}} & \multicolumn{1}{l}{\begin{tabular}[c]{@{}l@{}}4K\\ S2-S2\end{tabular}} & \multicolumn{1}{l}{\begin{tabular}[c]{@{}l@{}}4K\\ S2-S6\end{tabular}} & \multicolumn{1}{l}{\begin{tabular}[c]{@{}l@{}}1080p\\ S2-S6\end{tabular}} \\ \midrule
MeridianRoad & 332.78 & 107.05 & 102.71 & 1.02 & 1.02 & 1.16 & 0.07 & 0.07 & -0.2 \\
NocturneDance & 598.07 & 189.32 & 134.46 & 0.75 & 1.01 & 1.01 & 0.45 & -0.08 & -0.08 \\
NocturneRoom & 932.528 & 252.87 & 159.29 & 3.79 & 3.86 & 2.81 & -9.19 & -9.24 & -7.86 \\
SparksWelding & 1065.35 & 249.94 & 179.91 & 1.59 & 2.25 & 1.56 & -1.47 & -1.81 & -1.39 \\ \bottomrule
\end{tabular}
\end{table}

\subsection{Multiple Frame Types \texorpdfstring{$\lambda$}{lambda} Optimisation Performance}
\label{expt-result:multi_frame_results}
One of the objectives of this work is to compare different combinations of frame-type dependent $\lambda$ optimisations with the ultimate aim to find the best $\lambda$ optimisation strategy. First, we considered 4 modes, as in our previous work\cite{icip2022paper}. Particularly,  we set $\lambda = k\lambda_o$ for some grouped frame types, and kept $\lambda = \lambda_o$ for all other frame types. These four modes are: i) {\em All Frames}, which refers to the global $\lambda$ tuning method  where we set the same multiplier $k$ for all frames, as previously proposed by Ringis et al.~\cite{SPIERingis}, ii) {\em KF}, which refers to the scenario where we set $\lambda = k\lambda_o$ tune for KF, iii) {\em GF-ARF}, which refers to $\lambda$ optimisation for GF and ARF frames, iv) {\em KF-GF-ARF}, which refers to optimising $\lambda$ for KF, GF, and ARF frames. 

In addition to the above four modes, we introduced a new search method {\em Powell [KF-GF/ARF]}, which is a multidimensional joint search, where we deploy two $k$ values, with $k_1$ for {\em KF} and $k_2$  for {\em GF} and {\em ARF} frames. Thus, $\lambda_{KF} = k_1\lambda_o$ and $\lambda_{GF/ARF} = k_2\lambda_o$, and for others we use the default $\lambda$ value.

\begin{figure}[ht]
     \centering
     \begin{subfigure}[b]{0.3\textwidth}
         \centering
         \includegraphics[width=.8\textwidth]{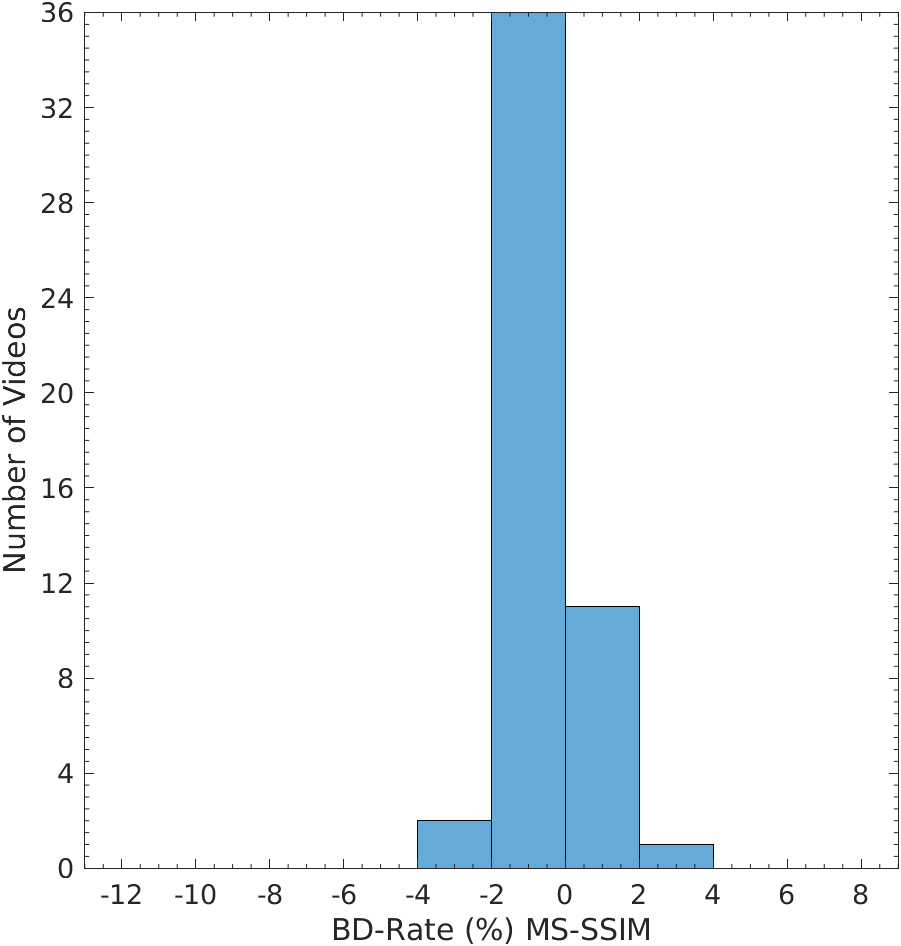}
         \caption{All Frames}
         \label{fig:brent_classic}
     \end{subfigure}
     \hfill
     \begin{subfigure}[b]{0.3\textwidth}
         \centering
         \includegraphics[width=.8\textwidth]{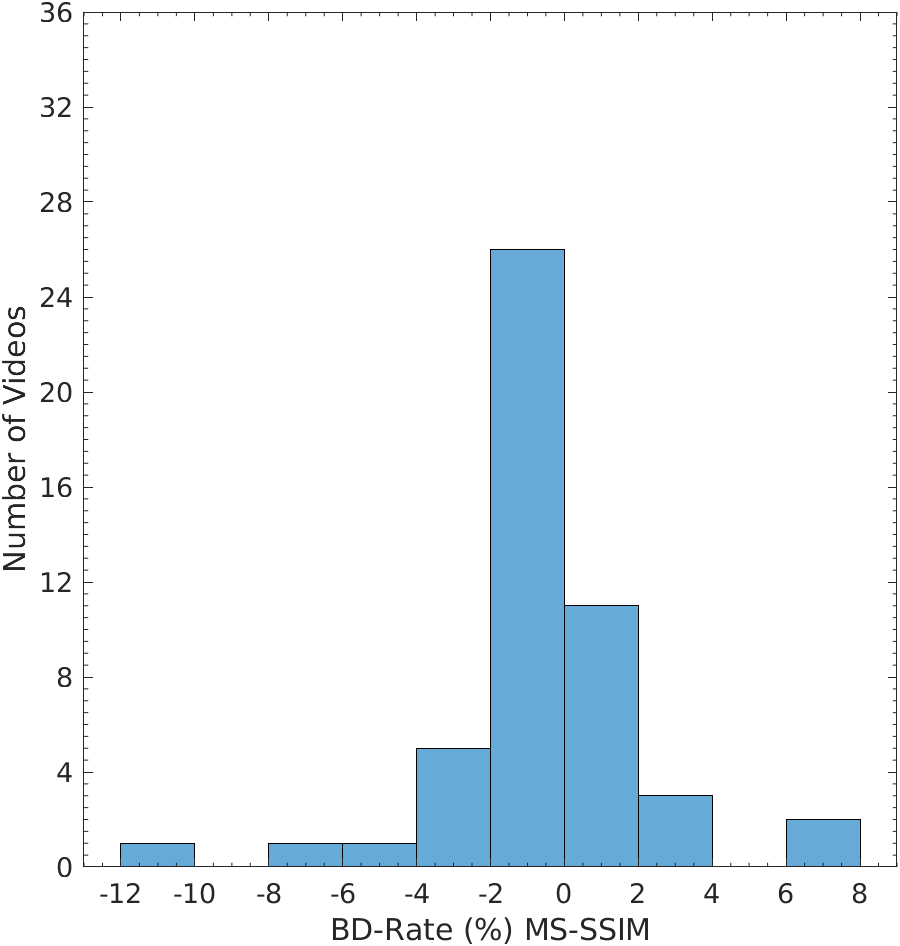}
         \caption{KF }
         \label{fig:brent_k1}
     \end{subfigure}
     \hfill
     \begin{subfigure}[b]{0.3\textwidth}
         \centering
         \includegraphics[width=.8\textwidth]{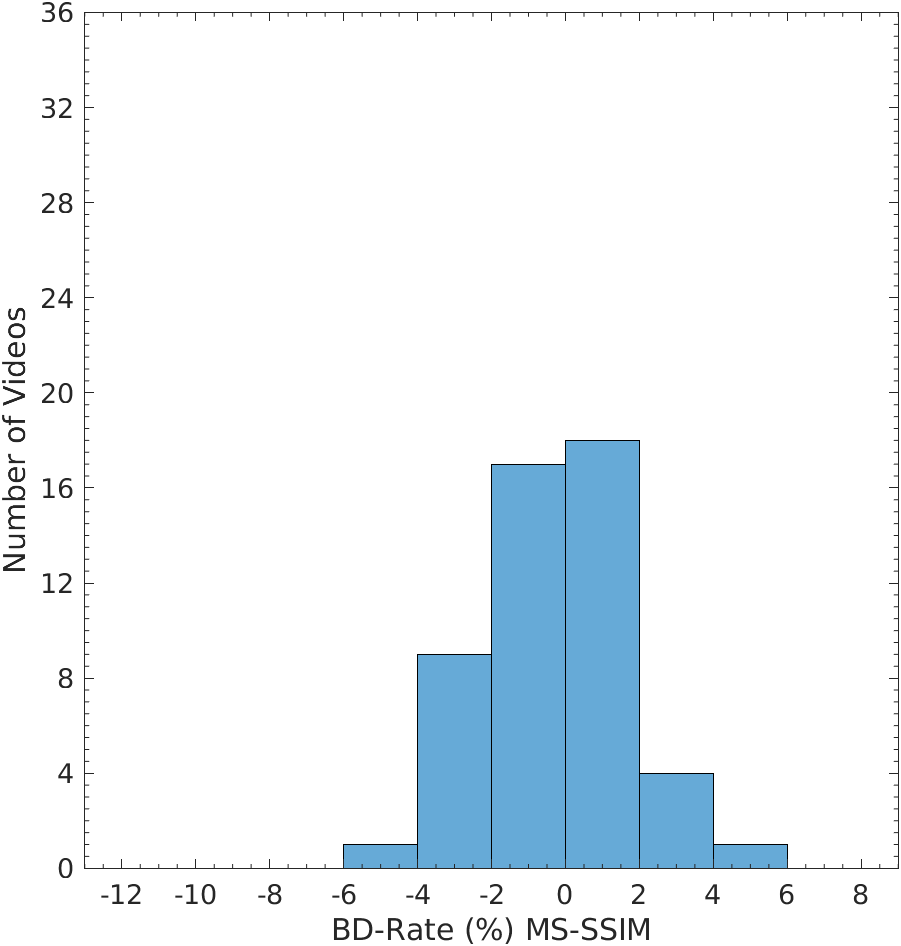}
         \caption{GF, ARF  }
         \label{fig:brent_k2}
     \end{subfigure}
     \hfill
 
     \begin{subfigure}[b]{0.3\textwidth}
         \centering
         \includegraphics[width=.8\textwidth]{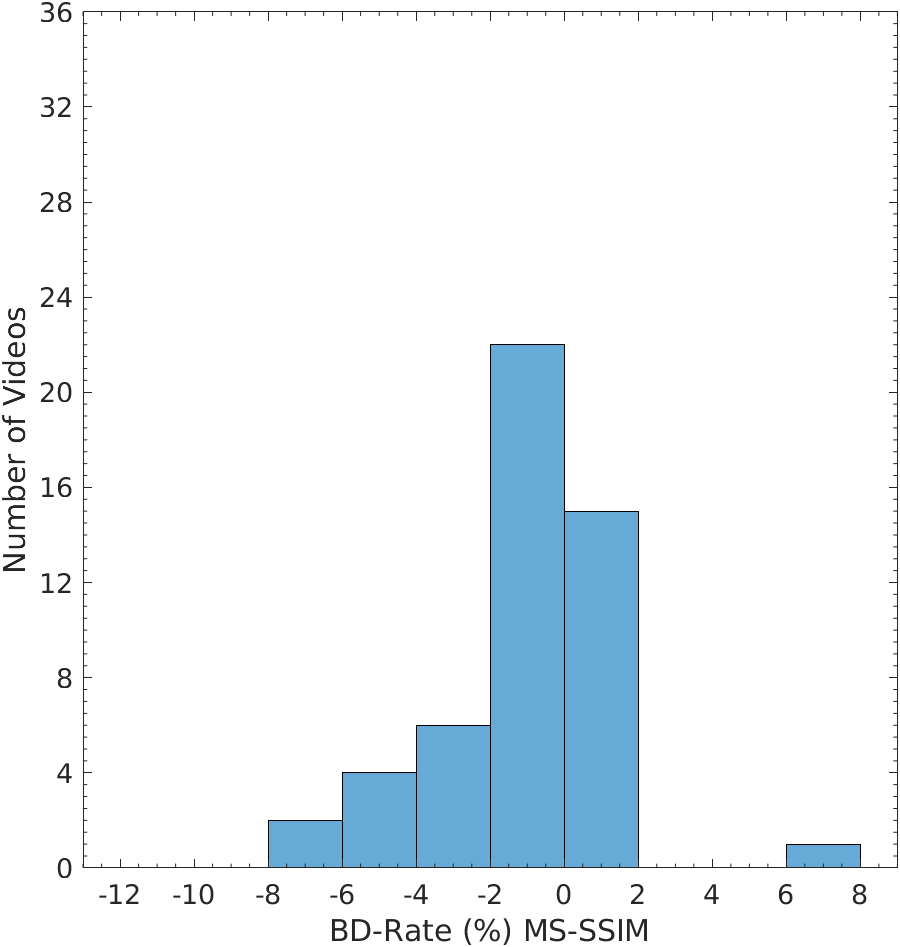}
         \caption{KF, GF, ARF }
         \label{fig:brent_t5}
     \end{subfigure}
    \hspace*{0.05\textwidth}%
     \begin{subfigure}[b]{0.3\textwidth}
         \centering
         \includegraphics[width=.8\textwidth]{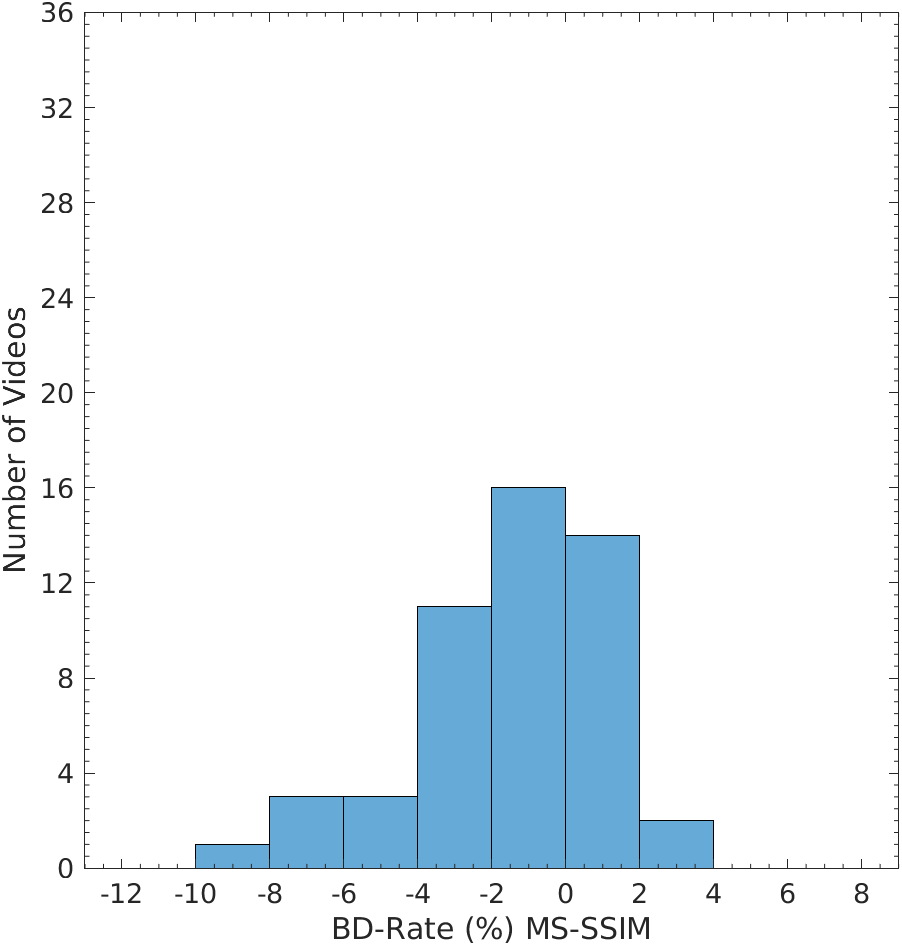}
         \caption{K1 = KF, K2 = GF/ARF}
         \label{fig:powell_k1k2}
     \end{subfigure}
         \hfill
         
         \vspace{-.5em}
        \caption{Histogram distribution of BD-rate(\%) MS-SSIM for various frame-level tuning methods. }
        \label{fig:histogram_dist}
\end{figure}

Table \ref{tab:full-gains-table} reports the BD-rate (MS-SSIM \%), MS-SSIM (dB), VMAF~\cite{vmafpaper} gains for these different optimisation strategies for the whole corpus which is divided into 7 different shots groups (see Table \ref{tab:seq-desc}). The underlined result shows the highest gains in terms of BD-rate that the proposed tuning brings compared to (4K S2). The results presented are averaged on a shot group basis. Also, the minimum  and average BD-rate in the shot group are recorded. Another perspective of the resulting BD-rates from all tested clips is illustrated in the histograms of the BD-rate values in Figure \ref{fig:histogram_dist}.

First, we observe that the new values of $k$ are, on average, significantly different from $k=1$, hence verifying our initial hypothesis of a ``better'' $\lambda$ value. Inspecting the Brent optimiser results from both Table~\ref{tab:full-gains-table} and Figure~\ref{fig:histogram_dist}, it transpires that the \textit{KF-GF-ARF} method is achieving the best BD-rate gains on average (1.12\%). Analysing the results on per-shot basis, we were able to see significant improvements in BD-rates for certain shots. Best average gains recorded were for Sol-Levante, where the BD-rate improved from 0.79\% to 2.94\%, and this was followed by Cosmos with an improvement from 1.44\% to 2.74\%. 

Another important observation is that the multidimensional search method {\em Powell [KF, GF/ARF]}, appears to be the overall top performer, with average BD-rate gains of 1.63\%. The histogram in Figure~\ref{fig:powell_k1k2} also shows that the improvement is consistent for the majority of the videos. This also evident in terms of bitrate, as we achieve a bitrate reduction of  0.64\% with \textit{KF, GF, ARF } and 4.25\% with {\em Powell [KF, GF/ARF]} at QP39. We also observed that for this improved bitrate savings, we have a minimal loss of MS-SSIM with $0.06$dB and $0.35$ for VMAF on average.

From Table~\ref{tab:full-gains-table}, we notice that clips within the same shot group respond differently. These deviations can be partly explained by variations in spatial and temporal information (See project page\footnotemark[\value{footnote}]). Analysing the distribution of results obtained with the Powell method, we observed that the higher BD-rate gains ( $>$ 4\%) were acquired for clips with high spatial information/complexity (5 clips). Videos with lower temporal information, exhibited higher gains. Videos with low temporal and low spatial complexity had no improvements. Different optimisation modes exhibited very different distributions, for instance, in the classic tuning method, the BD-rate improvements were noticed with clips with SI between 400-500 while for tuning mode of $\lambda$ for KF/GF gave large loss for the same range of videos.


\begin{table}[ht]
\caption{Per-frame-type $\lambda$-optimisation results for multiple combination of frame types. $k$ values are obtained using (1080p S6) proxy settings. BD-rates (BDR) are calculated using (4K S2) as an anchor. }
\resizebox{\textwidth}{!}{%
\label{tab:full-gains-table}
\begin{tabular}{@{}llrrrrrrrrr@{}}
\toprule
\textbf{Shot Group} & \textbf{Frame Type} & \multicolumn{1}{c}{\textbf{\begin{tabular}[c]{@{}r@{}}Avg. $k$\\ value\end{tabular}}} & \multicolumn{1}{c}{\textbf{\begin{tabular}[c]{@{}r@{}}Avg. \\ BDR (\%)\end{tabular}}} & \multicolumn{1}{c}{\textbf{\begin{tabular}[c]{@{}r@{}}Max.\\ BDR (\%)\end{tabular}}} & \multicolumn{1}{c}{\textbf{\begin{tabular}[c]{@{}r@{}}Min.\\ BDR (\%)\end{tabular}}} & \multicolumn{1}{c}{\textbf{\begin{tabular}[c]{@{}r@{}}Avg. \\ Iters\end{tabular}}} & \multicolumn{1}{c}{\textbf{\begin{tabular}[c]{@{}r@{}}Avg. \\ Bitrate.\\ Savings (\%)\end{tabular}}} & \multicolumn{1}{c}{\textbf{\begin{tabular}[c]{@{}r@{}}Avg. Q39 \\ Bitrate \\ Savings (\%)\end{tabular}}} & \multicolumn{1}{c}{\textbf{\begin{tabular}[c]{@{}r@{}}Avg. \\ MS-SSIM \\ Change (dB)\end{tabular}}} & \multicolumn{1}{c}{\textbf{\begin{tabular}[c]{@{}r@{}}Avg.\\ VMAF \\ Change\end{tabular}}} \\ \midrule
Cosmos & {\ul \textit{All Frames}} & {\ul \textit{1.49}} & {\ul \textit{-1.44}} & {\ul \textit{-2.91}} & {\ul \textit{-0.41}} & {\ul \textit{9.00}} & {\ul \textit{-15.71}} & {\ul \textit{-14.19}} & {\ul \textit{0.32}} & {\ul \textit{1.52}} \\
Meridian & All Frames & 1.07 & 0.12 & -0.29 & 0.36 & 9.71 & -5.32 & -3.45 & 0.04 & 0.24 \\
Nocturne & All Frames & 1.65 & -0.16 & -1.33 & 3.53 & 8.63 & -26.33 & -25.43 & 0.49 & 2.87 \\
Sol Levante & All Frames & 1.23 & -0.79 & -2.20 & 0.33 & 9.20 & -7.47 & -7.07 & 0.22 & 1.48 \\
Sparks & All Frames & 1.26 & -0.37 & -1.31 & 1.01 & 8.22 & -11.33 & -10.33 & 0.24 & 1.11 \\
SVT & All Frames & 0.97 & 0.04 & -0.36 & 1.04 & 8.83 & 3.12 & 3.04 & -0.14 & -0.24 \\
Cables 4K & All Frames & 1.56 & -0.36 & -1.05 & 0.16 & 10.00 & -16.69 & -15.81 & 0.44 & 1.74 \\ \midrule
Cosmos & KF  & 3.31 & -0.85 & -1.88 & 0.10 & 12.29 & -2.10 & -2.21 & 0.03 & 0.14 \\
Meridian & KF  & 4.00 & -0.37 & -10.45 & 7.86 & 11.00 & -2.03 & 1.41 & 0.04 & 0.19 \\
Nocturne & {\ul \textit{KF }} & {\ul \textit{5.08}} & {\ul \textit{-2.39}} & {\ul \textit{-6.98}} & {\ul \textit{-0.86}} & {\ul \textit{11.88}} & {\ul \textit{-4.28}} & {\ul \textit{-6.89}} & {\ul \textit{0.07}} & {\ul \textit{0.37}} \\
Sol Levante & KF  & 4.49 & -1.02 & -2.43 & 0.00 & 11.20 & -2.58 & -2.39 & 0.08 & 0.33 \\
Sparks & KF  & 2.54 & -0.44 & -2.75 & 2.21 & 11.33 & -0.44 & -0.78 & -0.01 & -0.03 \\
SVT & KF  & 6.20 & 0.36 & -0.59 & 3.01 & 10.50 & -1.83 & -1.74 & 0.13 & 0.38 \\
Cables 4K & KF  & 2.49 & 0.16 & -2.65 & 6.19 & 11.00 & 1.13 & 0.56 & -0.04 & -0.06 \\ \midrule
Cosmos & GF, ARF  & 2.00 & -1.63 & -3.88 & 0.00 & 8.00 & -2.69 & -2.53 & 0.04 & 0.17 \\
Meridian & GF, ARF  & 2.07 & 1.01 & -0.19 & 3.50 & 9.14 & -6.26 & -7.15 & 0.12 & 0.78 \\
Nocturne & GF, ARF  & 1.51 & 0.13 & -3.32 & 4.23 & 10.75 & 5.24 & 0.54 & 0.06 & 0.33 \\
Sol Levante & {\ul \textit{GF, ARF }} & {\ul \textit{2.71}} & {\ul \textit{-2.15}} & {\ul \textit{-3.42}} & {\ul \textit{-0.16}} & {\ul \textit{10.40}} & {\ul \textit{-4.87}} & {\ul \textit{-4.85}} & {\ul \textit{0.14}} & {\ul \textit{0.72}} \\
Sparks & GF, ARF  & 1.60 & 0.34 & -2.09 & 2.15 & 8.78 & -3.54 & -4.36 & 0.13 & 0.66 \\
SVT & GF, ARF  & 2.38 & -1.35 & -4.77 & 0.16 & 11.50 & -2.09 & -2.10 & 0.16 & 0.78 \\
Cables 4K & GF, ARF  & 1.66 & 0.40 & -0.59 & 2.23 & 9.50 & -5.21 & -6.15 & 0.20 & 0.63 \\ \midrule
Cosmos & KF, GF, ARF  & 3.36 & -2.74 & -6.89 & 0.55 & 10.86 & -5.58 & -5.66 & 0.10 & 0.43 \\
Meridian & KF, GF, ARF  & 0.96 & 1.37 & -0.53 & 6.92 & 10.86 & 1.01 & 4.41 & 0.02 & 0.02 \\
Nocturne & KF, GF, ARF  & 1.56 & -2.05 & -7.86 & 1.43 & 9.25 & -3.00 & -5.19 & 0.05 & 0.38 \\
Sol Levante & {\ul \textit{KF, GF, ARF }} & {\ul \textit{3.54}} & {\ul \textit{-2.94}} & {\ul \textit{-5.54}} & {\ul \textit{-0.25}} & {\ul \textit{11.20}} & {\ul \textit{-7.99}} & {\ul \textit{-7.48}} & {\ul \textit{0.24}} & {\ul \textit{1.27}} \\
Sparks & KF, GF, ARF  & 0.95 & -0.42 & -1.39 & 0.14 & 9.56 & 1.82 & 1.85 & -0.07 & -0.20 \\
SVT & KF, GF, ARF  & 0.91 & -0.97 & -3.81 & 0.12 & 10.67 & 4.81 & 4.55 & -0.27 & -0.45 \\
Cables 4K & KF, GF, ARF  & 0.93 & -0.09 & -0.81 & 0.37 & 9.75 & 1.56 & 1.49 & -0.06 & -0.17 \\ \midrule
Cosmos & Powell (KF, GF/ARF)& (3.99, 4.09) & -3.02 & -6.91 & 0.24 & 61.71 & -6.36 & -6.57 & 0.13 & 0.52 \\
Meridian & Powell (KF, GF/ARF)& (3.80, 1.12) & -1.13 & -9.30 & 3.92 & 40.71 & -4.42 & -5.45 & 0.03 & 0.12 \\
Nocturne & Powell (KF, GF/ARF)& (4.18, 1.27) & -2.80 & -7.84 & 0.00 & 52.63 & -3.64 & -5.73 & 0.07 & 0.43 \\
Sol Levante & {\ul \textit{Powell {[}KF,  GF/ARF{]}}} & {\ul \textit{(4.01, 3.61)}} & {\ul \textit{-3.43}} & {\ul \textit{-5.43}} & {\ul \textit{-2.28}} & {\ul \textit{100.40}} & {\ul \textit{-8.60}} & {\ul \textit{-8.22}} & {\ul \textit{0.25}} & {\ul \textit{1.32}} \\
Sparks & Powell (KF, GF/ARF)& (0.88, 1.58) & -0.33 & -2.17 & 1.68 & 60.33 & -0.15 & -0.82 & -0.01 & 0.11 \\
SVT & Powell (KF, GF/ARF)& (6.42, 0.91) & -0.76 & -3.62 & 0.01 & 47.00 & 0.88 & 0.83 & -0.09 & 0.03 \\
Cables 4K & Powell (KF, GF/ARF)& (2.00, 1.78) & -0.58 & -2.89 & 1.06 & 50.13 & -4.14 & -5.37 & 0.08 & 0.30 \\ \midrule
\multicolumn{1}{c}{\multirow{5}{*}{\textit{Average}}} & All Frames & 1.34 & -0.41 & -2.91 & 3.53 & 9.06 & -12.24 & -11.27 & 0.25 & 1.30 \\
\multicolumn{1}{c}{} & KF  & 3.88 & -0.66 & -10.45 & 7.86 & 11.34 & -1.64 & -1.71 & 0.04 & 0.17 \\
\multicolumn{1}{c}{} & GF, ARF  & 1.92 & -0.32 & -4.77 & 4.23 & 9.64 & -2.62 & -3.78 & 0.12 & 0.57 \\
\multicolumn{1}{c}{} & KF, GF, ARF  & 1.64 & -1.02 & -7.86 & 6.92 & 10.20 & -0.76 & -0.64 & -0.01 & 0.13 \\
\multicolumn{1}{c}{} & {\ul \textit{Powell (KF, GF/ARF)}} & \multicolumn{1}{l}{{\ul \textit{(3.27, 1.98)}}} & {\ul \textit{-1.63}} & {\ul \textit{-9.30}} & {\ul \textit{3.92}} & {\ul \textit{57.32}} & {\ul \textit{-3.46}} & {\ul \textit{-4.25}} & {\ul \textit{0.06}} & {\ul \textit{0.35}} \\ \bottomrule
\end{tabular}}%
\end{table}

\subsection{Convergence Speed of Optimisation Methods}
\label{expt-result:convergence}
The Powell method is computationally expensive, with a roughly twice computational cost compared to the Brent optimisation. Practically this means 203 hours on average over 118 hours per clip. Furthermore, the Brent search for optimal $k1$ and $k2$ determinations required an average of 15 hours, while Powell method's joint search for $(k1,k2)$ needed around 87 hours. The final optimisation step, i.e. the non-proxy encode at slower speed preset, took around 107 hours per clip on average for all three modes. 
It is interesting that for the Powell method and for most of the clips, we were able to achieve results very close to the final iteration by only using one iteration of the algorithm. 
With respect to the BD-RATE at iteration 1, the BD-rate at the final iteration (Powell method) was only on average 9\% different with a median of 2.4\%. Hence in practice, just one iteration of Powell is good enough and certainly for at least 50\% of the clips. 
Taking all the above into consideration, we can reduce the total computational cost by half just by reducing the iterations of Powell search. 

\subsection{HDR vs. SDR  \texorpdfstring{$\lambda$}{lambda} Optimisation}
\label{expt-result:hdr_vs_sdr}

It is educational to consider whether there is any difference in gains between equivalent HDR and SDR material. We therefore curated a subset of 39 clips from the current dataset, for which a SDR version of the sequence was released along with the HDR data from the content producer. 
Table \ref{tab:sdr_hdr_powell_k1k2} presents these SDR/HDR results for these particular subset sequences. The anchor for the BD-rate computations is the (4K S2) case. Overall, the average gains for HDR and SDR are comparable, with average BD-rate gains of 2.47\% for SDR vs. 1.89\% for HDR. Comparing directly the distributions of BD-rate between SDR and HDR, we notice that for 82\% of the clips we have very similar BD-rate gains ($\pm 1\%$).

We need to note that this comparison between HDR and SDR must however be nuanced by the fact that the same distortion metric (MS-SSIM) is applied in both HDR and SDR. Using the same distortion metric for both allows us to make a direct comparison, but MS-SSIM has not been designed for HDR. To the best of our knowledge, there is no objective quality metric available at the moment that could facilitate a direct comparison of SDR and HDR. Nevertheless the point is here that HDR optimisation and SDR optimisation is quite different. In particular Table \ref{tab:sdr_hdr_powell_k1k2} shows that the average estimated value of $\lambda$ is quite different for SDR and HDR.  

\begin{table}[ht]
\centering
\caption{$\lambda$ optimisation result for same set sequences represented in SDR and HDR domain, where the  $k$ values are obtained using (1080p S6) proxy settings. BD-rates (BDR) are calculated using (4K S2).}
\label{tab:sdr_hdr_powell_k1k2}
\resizebox{\textwidth}{!}{%
\begin{tabular}{@{}lllrrrrrrrr@{}}
\toprule
\textbf{\begin{tabular}[c]{@{}l@{}}Dynamic \\ Range\end{tabular}} & \textbf{Shot Group} & \textbf{\begin{tabular}[c]{@{}l@{}}Avg. $\lambda$ \\ value\end{tabular}} & \multicolumn{1}{l}{\textbf{\begin{tabular}[c]{@{}l@{}}Avg. \\ BDR (\%)\end{tabular}}} & \multicolumn{1}{l}{\textbf{\begin{tabular}[c]{@{}l@{}}Max.\\ BDR (\%)\end{tabular}}} & \multicolumn{1}{l}{\textbf{\begin{tabular}[c]{@{}l@{}}Min.\\ BDR (\%)\end{tabular}}} & \multicolumn{1}{l}{\textbf{\begin{tabular}[c]{@{}l@{}}Avg. \\ Iters\end{tabular}}} & \multicolumn{1}{l}{\textbf{\begin{tabular}[c]{@{}l@{}}Avg. \\ Bitrate \\ Savings (\%)\end{tabular}}} & \multicolumn{1}{l}{\textbf{\begin{tabular}[c]{@{}l@{}}Avg. Q39 \\ Bitrate\\ Savings (\%)\end{tabular}}} & \multicolumn{1}{l}{\textbf{\begin{tabular}[c]{@{}l@{}}Avg.\\ MS-SSIM \\ Change (dB)\end{tabular}}} & \multicolumn{1}{l}{\textbf{\begin{tabular}[c]{@{}l@{}}Avg. \\ VMAF\\ Change\end{tabular}}} \\ \midrule
 & {\ul \textit{Cosmos}} & {\ul \textit{{(}3.42, 4.41{)}}} & {\ul \textit{-3.26}} & {\ul \textit{-9.34}} & {\ul \textit{0.09}} & {\ul \textit{55.71}} & {\ul \textit{-5.98}} & {\ul \textit{-6.72}} & {\ul \textit{0.09}} & {\ul \textit{0.38}} \\
 & Meridian & {(}3.79, 0.89{)} & -2.78 & -9.9 & 4.18 & 59.33 & -3.32 & -5.52 & -0.01 & -0.13 \\
 & Nocturne & {(}4.11, 1.42{)} & -2.86 & -7.06 & 0.94 & 87.71 & -6.07 & -10.68 & 0.09 & 0.38 \\
\textbf{SDR} & Sol Levante & {(}5.36, 3.16{)} & -3.14 & -5.55 & -1.47 & 71 & -8.44 & -8.53 & 0.24 & 1.11 \\
 & Sparks & {(}0.82, 1.14{)} & -1.12 & -2.48 & 0.37 & 55 & 1.57 & 1.73 & -0.08 & -0.24 \\
 & SVT & {(}2.96, 1.60{)} & -2 & -10.69 & 0.14 & 53.17 & -1.41 & -1.42 & 0 & 0.33 \\ \cmidrule(l){2-11} 
 & \textit{Average} & {(}3.24, 2.07{)} & -2.47 & -10.69 & 4.18 & 63.44 & -3.65 & -4.93 & 0.04 & 0.26 \\ \midrule
 & Cosmos & {(}3.99, 4.09{)} & -3.02 & -6.91 & 0.24 & 61.71 & -6.36 & -6.57 & 0.13 & 0.52 \\
 & Meridian & {(}4.26, 1.15{)} & -1.32 & -9.3 & 3.92 & 43.33 & -5.15 & -6.35 & 0.04 & 0.13 \\
 & Nocturne & {(}3.60, 1.31{)} & -2.95 & -7.84 & 0 & 56.71 & -3.66 & -6.08 & 0.06 & 0.39 \\
\textbf{HDR} & {\ul \textit{Sol Levante}} & {\ul \textit{{(}4.01, 3.61{)}}} & {\ul \textit{-3.43}} & {\ul \textit{-5.43}} & {\ul \textit{-2.28}} & {\ul \textit{100.4}} & {\ul \textit{-8.6}} & {\ul \textit{-8.22}} & {\ul \textit{0.25}} & {\ul \textit{1.32}} \\
 & Sparks & {(}0.91, 1.61{)} & -0.28 & -2.17 & 1.68 & 59.5 & 0.11 & -0.45 & -0.02 & 0.07 \\
 & SVT & {(}6.42, 0.90{)} & -0.76 & -3.62 & 0.01 & 47 & 0.88 & 0.83 & -0.09 & 0.03 \\ \cmidrule(l){2-11} 
 & Average & {(}3.70, 2.08{)} & -1.89 & -9.3 & 3.92 & 60.23 & -3.53 & -4.27 & 0.06 & 0.37 \\ \bottomrule
\end{tabular}%
}
\end{table}



\section{Conclusion}
\label{conclusion}
We have presented a new method of per-clip optimisation based on the rate control $\lambda$ multiplier in AV1 for different frame types. The proposed method was tested on a 4K-HDR corpus of 50 videos. We reported improved average BD-rate gains of 1.6\% for the proposed per-frame-type per-clip optimisation compared to the 0.4\% for the global per-clip $\lambda$-optimisation. The proposed method showed improvements up to 3.4\% on average for certain shots compared to the method of global $\lambda$ tuning. The best improvements in BD-rate range from 2.1\% to 9.3\% with our proposed method. We have also showed that the computational complexity of the optimisation process can be mitigated by employing proxy settings and restricting the number of iterations used by the Powell optimiser. Thus, our proposed method of multivariable Powell-based optimiser gave the best improvements on average. Future work will focus on exploring the current implementation of deriving $\lambda$ from the quantiser along with more in-depth analysis of the quality aspect of the results including a subjective study. 


\acknowledgments 
 
This project is funded under Disruptive Technology Innovation Fund, Enterprise Ireland, Grant No DT-2019-0068, and  ADAPT-SFI Research Center, Ireland. 

\bibliography{report} 

\begin{thebibliography}{10}

\bibitem{av1paper}
Chen, Y., Murherjee, D., Han, J., Grange, A., Xu, Y., Liu, Z., Parker, S.,
  Chen, C., Su, H., Joshi, U., Chiang, C.-H., Wang, Y., Wilkins, P., Bankoski,
  J., Trudeau, L., Egge, N., Valin, J.-M., Davies, T., Midtskogen, S., Norkin,
  A., and de~Rivaz, P., ``An overview of core coding tools in the av1 video
  codec,'' in [{\em Picture Coding Symposium
  (PCS)}{\nolinebreak\hspace{0.1em}]},   41--45 (2018).

\bibitem{svtav1spie2021}
Wu, P.-H., Katsavounidis, I., Lei, Z., Ronca, D., Tmar, H., Abdelkafi, O.,
  Cheung, C., Amara, F.~B., and Kossentini, F., ``Towards much better svt-av1
  quality-cycles tradeoffs for vod applications,'' in [{\em Applications of
  Digital Image Processing XLIV}{\nolinebreak\hspace{0.1em}]},   {\bf 11842},
  236--256, SPIE (2021).

\bibitem{bdrate}
Bjontegaard, G., ``{Calculation of average PSNR differences between RD curves;
  VCEG-M33},'' in [{\em ITU-T SG16/Q6}{\nolinebreak\hspace{0.1em}]},  (2001).

\bibitem{netflix}
Aaron, A., Li, Z., Manohara, M., De~Cock, J., and Ronca, D., ``{Netflix
  Technology Blog - Per-Title Encode Optimization},'' (2019).
\newblock
  https://medium.com/netflix-techblog/per-title-encode-optimization-7e99442b62a2.

\bibitem{renznik_internetvideo2001}
Conklin, G.~J., Greenbaum, G.~S., Lillevold, K.~O., Lippman, A.~F., and Reznik,
  Y.~A., ``Video coding for streaming media delivery on the internet,'' {\em
  IEEE Trans. on Circuits and Systems for Video Technology}  (2001).

\bibitem{katsavounidis2018video}
Katsavounidis, I. and Guo, L., ``Video codec comparison using the dynamic
  optimizer framework,'' in [{\em Applications of Digital Image Processing
  XLI}{\nolinebreak\hspace{0.1em}]},   {\bf 10752}, SPIE (2018).

\bibitem{reznik2018abrstreaming}
Reznik, Y.~A., Lillevold, K.~O., Jagannath, A., Greer, J., and Corley, J.,
  ``Optimal design of encoding profiles for abr streaming,'' in [{\em
  Proceedings of the 23rd Packet Video Workshop}{\nolinebreak\hspace{0.1em}]},
   43--47 (2018).

\bibitem{TimmererSurvey}
{Bentaleb}, A., {Taani}, B., {Begen}, A.~C., {Timmerer}, C., and {Zimmermann},
  R., ``{A Survey on Bitrate Adaptation Schemes for Streaming Media Over
  HTTP},'' {\em IEEE Communications Surveys Tutorials}~{\bf 21}(1) (2019).

\bibitem{KatsenouOJSP2021}
Katsenou, A.~V., Sole, J., and Bull, D.~R., ``{Efficient Bitrate Ladder
  Construction for Content-Optimized Adaptive Video Streaming},'' {\em IEEE
  Open Journal of Signal Processing}~{\bf 2},  496--511 (2021).

\bibitem{pcs2021ringis}
Ringis, D.~J., Pitié, F., and Kokaram, A., ``Near optimal per-clip lagrangian
  multiplier prediction in hevc,'' in [{\em 2021 Picture Coding Symposium
  (PCS)}{\nolinebreak\hspace{0.1em}]},   1--5 (2021).

\bibitem{SPIERingis}
Ringis, D.~J., Piti{\'e}, F., and Kokaram, A., ``Per-clip adaptive lagrangian
  multiplier optimisation with low-resolution proxies,'' in [{\em Applications
  of Digital Image Processing XLIII}{\nolinebreak\hspace{0.1em}]},   {\bf
  11510},  115100E, International Society for Optics and Photonics (2020).

\bibitem{icip2022paper}
Vibhoothi, Piti\'e, F., and Kokaram, A., ``{Frame-type Sensitive RDO Control
  for Content-Adaptive-encoding},'' {\em ArxiV Preprint [Online]}~{\bf
  2206.11976} (2022).

\bibitem{itu_hdr}
ITU-R, R., ``{BT2100-2: image parameter values for high dynamic range
  television for use in production and international programme exchange},''
  (2018).

\bibitem{sullivan1998rate}
Sullivan, G.~J. and Wiegand, T., ``Rate-distortion optimization for video
  compression,'' {\em IEEE signal processing magazine}~{\bf 15}(6),  74--90
  (1998).

\bibitem{zhangbulllambdahevc}
Zhang, F. and Bull, D.~R., ``Rate-distortion optimization using adaptive
  lagrange multipliers,'' {\em IEEE Trans. on Circuits and Systems for Video
  Technology}~{\bf 29}(10),  3121--3131 (2019).

\bibitem{EIRingis}
Ringis, D.~J., Pitie, F., and Kokaram, A., ``Per clip lagrangian multiplier
  optimisation for ({HEVC}),'' {\em Electronic Imaging}~{\bf 2020}(12) (2020).

\bibitem{numericalmethods}
Flannery, B.~P., Press, W.~H., Teukolsky, S.~A., and Vetterling, W.,
  ``Numerical recipes in c,'' {\em Press Syndicate of the University of
  Cambridge, New York}~{\bf 24},  78 (1992).

\bibitem{hdr4kcomparision}
Pourazad, M.~T., Sung, T., Hu, H., Wang, S., Tohidypour, H.~R., Wang, Y.,
  Nasiopoulos, P., and Leung, V.~C., ``{Comparison of Emerging Video
  Compression Schemes for Efficient Transmission of 4K and 8K HDR Video},'' in
  [{\em IEEE International Mediterranean Conference on Communications and
  Networking}{\nolinebreak\hspace{0.1em}]},  (2021).

\bibitem{topiwala2021hdr}
Topiwala, P. and Dai, W., ``{HDR video coding for aerial videos with VVC and
  AV1},'' in [{\em Applications of Digital Image Processing
  XLIV}{\nolinebreak\hspace{0.1em}]},   {\bf 11842},  118420J, International
  Society for Optics and Photonics (2021).

\bibitem{barman_ugc_hdr}
Barman, N. and Martini, M.~G., ``User generated hdr gaming video streaming:
  dataset, codec comparison and challenges,'' {\em IEEE Trans. on Circuits and
  Systems for Video Technology}~{\bf abs/2103.02189} (2021).

\bibitem{zhou_lambda_hdr}
Zhou, M., Wei, X., Wang, S., Kwong, S., Fong, C.-K., Wong, P. H.~W., and Yuen,
  W. Y.~F., ``{Global Rate-Distortion Optimization-Based Rate Control for HEVC
  HDR Coding},'' {\em IEEE Trans. on Circuits and Systems for Video
  Technology}~{\bf 30}(12),  4648--4662 (2020).

\bibitem{hdrvdp2}
Mantiuk, R., Kim, K.~J., Rempel, A.~G., and Heidrich, W., ``{HDR-VDP-2: A
  calibrated visual metric for visibility and quality predictions in all
  luminance conditions},'' {\em ACM Trans. on Graphics} , ACM New York, NY, USA
  (2011).

\bibitem{msssimpaper}
Wang, Z., Simoncelli, E., and Bovik, A., ``Multiscale structural similarity for
  image quality assessment,'' in [{\em 37th Asilomar Conference on Signals,
  Systems Computers, 2003}{\nolinebreak\hspace{0.1em}]},   {\bf 2},  1398--1402
  Vol.2 (2003).

\bibitem{ping_spie_proxy}
Wu, P.-H., Kondratenko, V., and Katsavounidis, I., ``Fast encoding parameter
  selection for convex hull video encoding,'' in [{\em Applications of Digital
  Image Processing XLIII}{\nolinebreak\hspace{0.1em}]},   {\bf 11510},
  181--194, SPIE (2020).

\bibitem{ping_pcs_proxy}
Wu, P.-H., Kondratenko, V., Chaudhari, G., and Katsavounidis, I., ``Encoding
  parameters prediction for convex hull video encoding,'' in [{\em 2021 Picture
  Coding Symposium (PCS)}{\nolinebreak\hspace{0.1em}]},   1--5, IEEE (2021).

\bibitem{simplex_paper}
Gao, F. and Han, L., ``Implementing the nelder-mead simplex algorithm with
  adaptive parameters,'' {\em Computational Optimization and Applications}~{\bf
  51}(1),  259--277 (2012).

\bibitem{nocedal2006conjugate}
Nocedal, J. and Wright, S.~J., ``Conjugate gradient methods,'' {\em Numerical
  optimization} ,  101--134 (2006).

\bibitem{powellmethodpaper}
Powell, M.~J., ``An efficient method for finding the minimum of a function of
  several variables without calculating derivatives,'' {\em The computer
  journal}~{\bf 7}(2),  155--162 (1964).

\bibitem{aomctc}
Xin, Z., Zhijun(Ryan), L., Andrey, N., Thomas, D., and Alexis, T., ``{AOM
  Common Test Conditions v2.0},'' {\em Alliance for Open Media, Codec Working
  Group Output Document}~{\bf CWG/B075o} (2021).
\newblock \url{https://aomedia.org/docs/CWG-B075o_AV2_CTC_v2.pdf}.

\bibitem{hdrtools}
ITU-T and ISO/IEC, ``{HDRTools pacakge [Online]},'' (2015).
\newblock Available: \url{https: gitlab.com/standards/HDRTools}.

\bibitem{itup910}
Recommendation, I., ``{Subjective video quality assessment methods for
  multimedia applications},'' (2021).
\newblock ITU-T P.910.

\bibitem{netflixopencontent}
Netflix, ``Netflix open content.''
\newblock Available: \url{https://opencontent.netflix.com/}.

\bibitem{svt2022}
Josef, A., Olof, L., Marcus, L., and Fredrik, L., ``{SVT OpenContent Video Test
  Suite 2022– Natural Complexity},'' (2022).
\newblock Sveriges Television AB, https://www.svt.se/open/en/content/.

\bibitem{cablelabs}
Cable Television~Laboratories, I., ``{4K Video Set},'' (2014).
\newblock https://www.cablelabs.com/4k.

\bibitem{av1:adaptivepred}
Liu, Z., Mukherjee, D., Lin, W.-T., Wilkins, P., Han, J., and Xu, Y.,
  ``Adaptive multi-reference prediction using a symmetric framework,'' {\em
  Electronic Imaging}~{\bf 2017}(2),  65--72 (2017).

\bibitem{av1compound}
Chen, C., Han, J., and Xu, Y., ``A hybrid weighted compound motion compensated
  prediction for video compression,'' in [{\em Picture Coding Symposium
  (PCS)}{\nolinebreak\hspace{0.1em}]},   223--227 (2018).

\bibitem{libvmafurl}
Netflix, ``{VMAF - Video Multi-Method Assessment Fusion},'' (2016).
\newblock https://github.com/Netflix/vmaf.

\bibitem{awcy}
Xiph~Org, F., ``{Are We Compressed Yet, AWCY Source[Online]},'' (2015).
\newblock \url{https://github.com/xiph/awcy}.

\bibitem{vmafpaper}
Lin, J.~Y., Liu, T.-J., Wu, E. C.-H., and Kuo, C.-C.~J., ``A fusion-based video
  quality assessment {(FVQA)} index,'' in [{\em Signal and Information
  Processing Association Annual Summit and Conference
  (APSIPA)}{\nolinebreak\hspace{0.1em}]},  (2014).

\end{thebibliography}
\bibliographystyle{spiebib} 

\end{document}